\documentclass[letterpaper,twocolumn,showpacs,preprintnumbers,amsmath,amssymb,floatfix,superscriptaddress,pre]{revtex4}

\usepackage{graphicx}  
\usepackage{dcolumn}   
\usepackage{bm}        
\usepackage{epsfig}

\begin{document}

\title{Noise-dependent stability of the synchronized state \\
in a coupled system of active rotators}

\author{Sebastian F.~Brandt}
\email{sbrandt@physics.wustl.edu}
\affiliation{
Department of Physics, Campus Box 1105, Washington University in St.~Louis, MO 63130-4899, USA}

\author{Axel Pelster}
\email{axel.pelster@uni-duisburg-essen.de}
\affiliation{%
Universit{\"a}t Duisburg-Essen, Campus Duisburg, Fachbereich Physik, Lotharstra{\ss}e 1, 47048 
Duisburg, Germany}%

\author{Ralf Wessel}
\email{rw@physics.wustl.edu}
\affiliation{
Department of Physics, Campus Box 1105, Washington University in St.~Louis, MO 63130-4899, USA}

\date{February 07, 2008}
\begin{abstract}
We consider a Kuramoto model for the dynamics of an excitable system
consisting of two coupled active rotators. Depending on both the coupling
strength and the noise, the two rotators can be in a synchronized or
desynchronized state. The synchronized state of the system is most stable
for intermediate noise intensity in the sense that the coupling strength
required to desynchronize the system is maximal at this noise level. We
evaluate the phase boundary between synchronized and desynchronized
states through numerical and analytical calculations.
\end{abstract}
\maketitle
%
\section{Introduction}
Networks of coupled nonlinear oscillators provide useful model systems for the study of a variety of phenomena in physics and biology \cite{Heagy}. Among many others, examples from physics include solid-state lasers \cite{Roy} and coupled Josephson junctions \cite{Ustinov,Wiesenfeld1}.  In biology, the central nervous system can be described as a complex network of oscillators \cite{Amit}, and cultured networks of heart cells are examples of biological structures with strong nearest-neighbor coupling \cite{Soen}.  In particular, the emergence of synchrony in such networks \cite{Pikovsky,Strogatz} has received increased attention in recent years.

Disorder and noise in physical systems usually tend to destroy spatial and temporal regularity. However, in nonlinear systems, often the opposite effect is found and intrinsically noisy processes, such as thermal fluctuations or mechanically randomized scattering, lead to surprisingly ordered patterns \cite{Shinbrot}.  For instance, arrays of coupled oscillators can be synchronized by randomizing the phases of their driving forces \cite{Brandt1,Chacon}. Synchronization in these systems is caused by the interactions between the elements and results in the emergence of collective modes. It has been shown to be a fundamental mechanism of self-organization and structure formation in systems of coupled oscillators \cite{Zaks1}. Biological systems of neurons are subject to different sources of noise, such as synaptic noise \cite{Calvin} or channel noise \cite{White}. In particular, sensory  neurons are notoriously noisy. Therefore, the question arises how stochastic influences affect the functioning of biological systems. Especially interesting are scenarios in which noise enhances performance. In the case of stochastic resonance \cite{Hanggi}, e.g., noise can improve the ability of a system to transfer information reliably, and the presence of this phenomenon in neural systems has been investigated \cite{Douglass,Wiesenfeld2}. Furthermore, numerous studies have addressed the effect of noise on the dynamics of limit cycle systems \cite{Treutlein,Kurrer1,Kurrer2,Kurrer3,Zaks1,Zaks2,Lindner}.

Small neural circuits composed of two or three neurons form the basic feedback mechanisms involved in the regulation of neural activity \cite{Milton}. They can display oscillatory activity \cite{Brandt2,Brandt3} and serve as central pattern generators involved in motor control \cite{Rabinovich}.  Here, we consider a system of two limit cycle oscillators with repulsive coupling. We investigate the influence of the noise and the coupling strength on the dynamics of the system. We distinguish between two different classes of dynamics, a synchronized state, in which the joint probability density of the oscillator phases is characterized by a single-hump shape, and a desynchronized state. The single-hump shaped distribution of the oscillator phases has been modeled by a Gaussian distribution \cite{vanKampen,Zaks1}, and systems consisting of a large number of oscillators were analyzed by examining the resulting dynamics for the mean of the oscillator phases \cite{Kurrer2}. In contrast, the simplicity of our two oscillator system allows us to obtain the stationary probability density function for the full system both numerically and analytically. We show that the probability distribution of the oscillator phases has the single-hump shape only for weak coupling, whereas it deviates from this shape for strong coupling. We evaluate the coupling strength at which the transition between the two forms of the probability distribution occurs as a function of the noise intensity.

In Sect.\ \ref{Sect2}, we introduce the Kuramoto model for excitable systems. Under the influence of noise, the dynamics of the limit cycle oscillators are described by a stochastic differential equation (SDE), and we state the Fokker-Planck equation for the system. In Sect.\ \ref{Sect3}, we consider a single active rotator driven by noise and derive its mean angular frequency from the stationary solution to the Fokker-Planck equation. We compare our analytical results with Monte-Carlo simulations of the corresponding SDE. In Sect.\ \ref{Sect4}, we consider two coupled deterministic rotators and perform a bifurcation analysis of the system. We show that the system possesses a fixed point that is stable for small coupling strengths but looses its stability when the coupling is increased. For some range of the coupling strength, the stable fixed point and a stable limit cycle coexist. In Sect.\ \ref{Sect5}, we consider two coupled active rotators under uncorrelated stochastic influences. In Sect.\ \ref{Sect5a}, we solve the Fokker-Planck equation of the system numerically and show that the shape of the probability distribution undergoes a characteristic change, corresponding to the transition from a synchronized to a desynchronized state, as coupling is increased. We evaluate the boundary between the synchronous and the asynchronous regime through a Fourier expansion approach in Sect.\ \ref{Sect5b}. A summary concludes the paper in Sect.\ \ref{Sect6}.

\section{Excitable Systems and the Kuramoto Model} \label{Sect2}

Neurons can display a wide range of behavior to different stimuli and numerous models exist to describe neuronal dynamics.  A common feature of both biological and model neurons is that sufficiently strong input causes them to fire periodically; the neuron displays oscillatory activity.  For subthreshold inputs, on the other hand, the neuron is quiescent. When a subthreshold input is combined with a noisy input, however, the neuron will be pushed above threshold from time to time and fire spikes in a stochastic manner.  In this regime, the neuron acts as an excitable element. In general, an excitable system possesses a stable equilibrium point from which it can temporarily depart by a large excursion through its phase space when it receives a stimulus of sufficient strength \cite{Zaks2}. Besides neurons, chemical reactions, lasers, models of blood clotting, and cardiac tissues all display excitable dynamics \cite{Sakurai,Wunsche,Lobanova,Panfilov,Koch}. Pulse propagation, spiral waves, spatial and temporal chaos, and synchronization have been studied in these systems \cite{Murray,Mikhailov,Chay,Hu}.
 
The phase dynamics of an active rotator without interaction and random forces can be described by the model developed by Kuramoto and coworkers \cite{Kuramoto86,Kuramoto88}:
\begin{eqnarray}
\dot{\phi}(t) = \omega - a \sin \phi(t)\, .
\end{eqnarray}
To obtain the case of the excitable system with one stationary point, one chooses the parameter $a>\omega$. When we have $n$ coupled identical oscillators, subject to stochastic influences, the model is described by the Langevin equation \cite{Lindner}
\begin{eqnarray}
\dot{\phi}_i(t) = \omega - a \sin \phi_i(t) - \sum_{j=1}^n W_{ij} (\phi_j - \phi_i) + \eta_i(t)\, \label{dyn}.
\end{eqnarray}
Here, we take the $\eta_i$ to be uncorrelated Gaussian white noise, i.e., $\langle \eta_i(t) \rangle = 0$, $\langle \eta_i(t_1) \eta_j(t_2) \rangle = 2 \sigma \delta(t_1 - t_2)\delta_{ij}$. We will concentrate on the simplest case, namely that the coupling functions $W_{ij}$ are $\sin$-functions multiplied by a coupling constant $w_{ij}$, i.e., $W_{ij}(\phi) = w_{ij} \sin \phi$. Then, the dynamical evolution of the system's probability density function $P({\boldsymbol \phi}, t)$ is described by the Fokker-Planck equation
\begin{eqnarray}
\frac{ \partial}{\partial t} P( {\boldsymbol \phi}, t)  & = & - \sum_{i = 1}^n\frac{\partial}{\partial \phi_i}\left[D_i({\boldsymbol \phi}) P({\boldsymbol \phi}, t) \right] \label{FPe} \\
&& + \sum_{i = 1}^n \sum_{j = 1}^n\frac{\partial^2}{\partial \phi_i \partial \phi_j} \left[D_{ij}({\boldsymbol \phi}) P({\boldsymbol \phi},t) \right]\, , \nonumber
\end{eqnarray}
where in our case the drift terms read
\begin{eqnarray}
D_i({\boldsymbol \phi}) = \omega - a \sin \phi_i - \sum_{j=1}^n w_{ij} \sin(\phi_j - \phi_i) \label{drift}
\end{eqnarray}
and the diffusion terms are given by
\begin{eqnarray}
D_{ij}({\boldsymbol \phi}) = \delta_{ij} \sigma \, . \label{diff}
\end{eqnarray}
Since the angle variables $\phi_i$ describe the phases of the oscillators, the probability density function must satisfy the periodic boundary conditions
\begin{eqnarray}
P(\phi_1,\,  \cdots,\, \phi_i  = 0, \, \cdots,\, \phi_n, t) \label{perc} \hspace{35mm}\\
= P(\phi_1,\,  \cdots,\, \phi_i  = 2 \pi, \, \cdots,\, \phi_n, t)\, , \quad i = 1,\, \cdots,\, n\, . \nonumber
\end{eqnarray}
Furthermore, the normalization condition for the probability density reads
\begin{eqnarray}
\int_0^{2\pi} d \phi_1 \cdots \int_0^{2\pi} d \phi_n  P({\boldsymbol \phi},t) = 1\, . \label{normc} 
\end{eqnarray}
\section{Single-Rotator System} \label{Sect3}
We first exam a single rotator subject to a noisy input and, following Ref.~\cite{Risken}, calculate the mean frequency of oscillations as a function of the noise level. In this case, the Fokker-Planck equation (\ref{FPe}) reads
\begin{eqnarray}
\frac{\partial}{\partial t} P(\phi, t)= - \frac{\partial}{\partial \phi}\left[D(\phi) P(\phi, t)\right] + \sigma \frac{\partial^2}{\partial \phi^2}P(\phi, t)\, ,
\end{eqnarray}
with
\begin{eqnarray}
D(\phi) = \omega - a \sin \phi \, .
\end{eqnarray}
We can thus write the drift term as the negative gradient of a potential, $D = - \partial V / \partial \phi$, with the potential given by 
\begin{eqnarray}
V(\phi) = - \omega \phi - a \cos \phi + c \, . \label{Vdef}
\end{eqnarray}
Introducing the probability current
\begin{eqnarray}
S(\phi, t) = D(\phi)P(\phi, t) - \sigma \frac{\partial}{\partial \phi} P(\phi, t)\, ,
\end{eqnarray}
the Fokker-Planck equation takes the form of a continuity equation,
\begin{eqnarray}
\frac{\partial}{\partial t} P(\phi, t) + \frac{\partial}{\partial \phi} S(\phi, t) = 0 \, . \label{conteq}
\end{eqnarray}
We now look for a stationary solution of the form $P(\phi, t) = P(\phi)$, $S(\phi, t) = S(\phi)$. In this case, we conclude from (\ref{conteq}) that the derivative of the probability current with respect to $\phi$ must vanish, and we have to solve
\begin{eqnarray}
S = D(\phi)P(\phi) - \sigma \frac{\partial}{\partial \phi} P(\phi) \, . \label{ODESconst}
\end{eqnarray}
The constant probability current $S$ is related to the mean drift velocity, i.e., the mean angular frequency of the active rotator system according to $\bar{\omega} = 2 \pi S$.
The solution to the ordinary differential equation (\ref{ODESconst}) is given by
\begin{eqnarray}
P(\phi) = C e^{-\frac{V(\phi)}{\sigma}} - \frac{S}{\sigma} \int_0^{\phi} d \phi' e^{\frac{V(\phi') - V(\phi)}{\sigma}}\, . \label{ODEsol}
\end{eqnarray}
The integration constant in (\ref{Vdef}) can thus be absorbed into the constant $C$ in (\ref{ODEsol}), and the two free constants $S$ and $C$ are determined by the periodicity and normalization conditions (\ref{perc}) and (\ref{normc}). These two conditions can be written in matrix form as
\begin{eqnarray}
\left( \begin{array}{cc}
\int_0^{2\pi} d \phi \hspace{1mm} e^{- \frac{V(\phi)}{\sigma}} & \int_0^{2 \pi}d \phi \int_0^{\phi} d \phi' e^{\frac{V(\phi') - V(\phi)}{\sigma}}  \\
e^{-\frac{V(2 \pi)}{\sigma}}-e^{-\frac{V(0)}{\sigma}} & \int_0^{2 \pi}d \phi \hspace{1mm} e^{\frac{V(\phi) - V(2 \pi)}{\sigma}}  \end{array} \right) 
\left( \begin{array}{c}
C \\
-\frac{S}{\sigma}
\end{array}
\right)
\nonumber 
&&
\hspace{-5mm}
\\
&\hspace{-44mm}=& \hspace{-22mm}
\left( \begin{array}{c}
1 \\
0
\end{array}
\right) .
\end{eqnarray}
Denoting the determinant of the $2\times 2$ matrix in the last expression as $\det$, the constants $C$ and $S$ are given by
\begin{eqnarray}
C &=& \frac{e^{-\frac{V(2\pi)}{\sigma}}}{\det} \int_0^{2 \pi}d \phi \hspace{1mm} e^{\frac{V(\phi)}{\sigma}}\, , \\
S &=& \frac{\sigma}{\det}\left[e^{-\frac{V(2\pi)}{\sigma}}-e^{-\frac{V(0)}{\sigma}}\right]\, .
\end{eqnarray}
Specializing to the potential of the active rotator (\ref{Vdef}), we obtain  
\begin{figure}[t]
  \begin{center}
    \epsfig{file=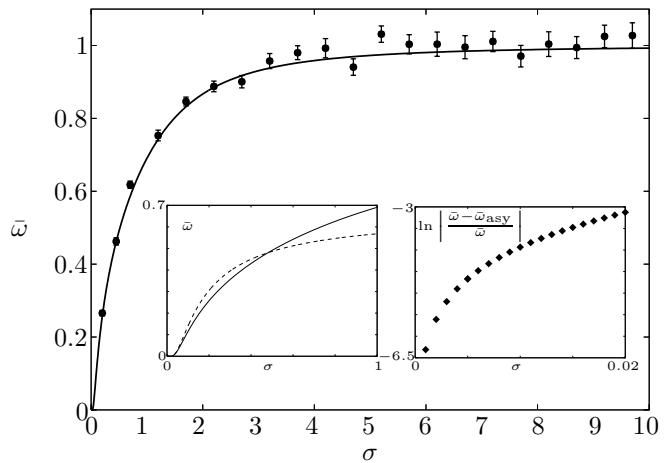,width=\columnwidth}        
\caption{Average angular frequency of the single-rotator as a function of the noise intensity. The solid line shows the result (\ref{freqint}). The dots represent results from Monte-Carlo simulations (mean $\pm$ standard error of the mean) of the Langevin equation (\ref{dyn}). For each value of the noise intensity, forty runs where simulated up to $T = 400$. The first inset shows a comparison between the asymptotic expansion (\ref{omasy}, dashed line) and numerical evaluations of the expression (\ref{freqint}, solid lines) for small noise. The diamonds in the second inset show the logarithm of the relative deviation between the result (\ref{freqint}) and its asymptotic approximation (\ref{omasy}). Parameters are: $\omega = 1$, $a=1.2$.}
    \label{fig1}
  \end{center}
\end{figure} 
\begin{figure*}[t]
  \begin{center}
    \epsfig{file=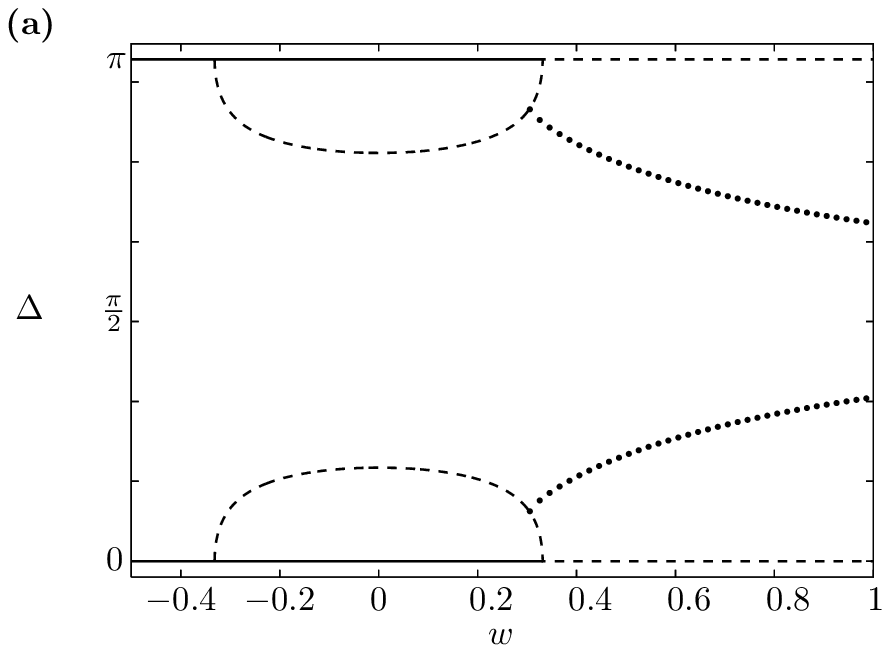,width=\columnwidth} 
    \hfill
    \epsfig{file=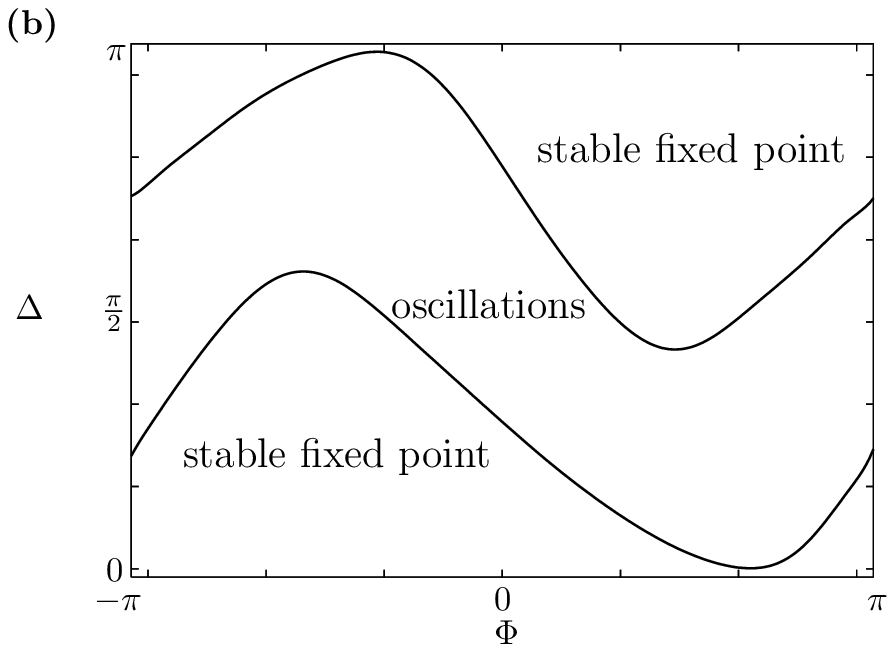,width=\columnwidth}
    \caption{Stable and unstable fixed points and oscillations in the deterministic two-rotator system. {\bf (a)} shows the bifurcation diagram with stable (solid lines) and unstable (dashed lines) fixed points of the system (\ref{2oscdet}) for the choice of parameters $\omega = 1$, $a=1.2$, $w_{12}=w_{21}=w$.  Dots indicate the minimum and maximum values of oscillations in the value of $\Delta$ that result for the initial conditions $\Phi=0$, $\Delta=\pi/2$. {\bf (b)} depicts for $w=0.308$ the boundaries between the regions in the space of initial conditions for which the system converges to the limit cycle or the stable fixed point.}
    \label{fig2}
  \end{center}
\end{figure*}
\begin{eqnarray}
\bar{\omega} = \frac{2 \pi \sigma\ \left(1 - e^{- \frac{2 \pi \omega}{\sigma}}\right)}
{\int_0^{2 \pi} d \phi'  \hspace{1mm} e^{-\frac{\omega}{\sigma}\phi' } \int_0^{2 \pi} d \phi \hspace{1mm} e^{ \frac{a}{\sigma} \left[ \cos(\phi + \phi') - \cos \phi \right] } } \, . \label{freqint}
\end{eqnarray}
Note that in the limit $\sigma \rightarrow \infty$ the integrand in the denominator approaches one, and $\bar{\omega}$ converges to $\omega$. To obtain the leading order behavior of $\bar{\omega}$ in the limit of small noise, we approximate the denominator using Laplace's method described in Ref.~\cite{Bender}. According to Laplace's method the asymptotic behavior of the integral
\begin{eqnarray}
I(x) = \int_a^b dt f(t)e^{x g(t)}
\end{eqnarray}
as $x \rightarrow \infty $ is given by
\begin{eqnarray}
I(x) \sim \frac{ \sqrt{2 \pi} f(c) e^{x g(c)}}{\sqrt{-x g''(c)}} 
\, . \label{Lap}
\end{eqnarray}
Here, it is assumed that $g(t)$ has a maximum at $t = c$ with $a \leq c \leq b$ and that $f(c) \neq 0$ and $g''(c) < 0$. We first apply Laplace's method to the inner integral in the denominator of (\ref{freqint}), which we denote as $I(\sigma)$. The function $a[\cos(\phi + \phi') - \cos \phi]$ has a maximum inside the interval $0 \leq \phi \leq 2 \pi$ at
\begin{eqnarray}
\phi_0 = \pi + \arctan \frac{\sin \phi'}{1 - \cos \phi'} \, .
\end{eqnarray}
Using (\ref{Lap}) we thus obtain for $\sigma \rightarrow 0$
\begin{eqnarray}
I(\sigma) \sim \sqrt{\frac{2 \pi \sigma}{a}} \int_0^{2 \pi} d \phi' \hspace{1mm} \frac{e^{\frac{a}{\sigma}[\cos(\phi_0 + \phi') - \cos \phi_0] - \frac{\omega}{\sigma} \phi'}}{\sqrt{\cos(\phi_0 + \phi') - \cos \phi_0}}\, . \label{intmed1}
\end{eqnarray}
The argument of the exponential function in the last identity can be simplified to
\begin{eqnarray}
\frac{a - \cos \phi'}{\sqrt{\sin^2 \frac{\phi'}{2}}} - \omega \phi'\, ,
\end{eqnarray}
whose maximum within the interval $0\leq \phi' \leq 2 \pi$ is at
\begin{eqnarray}
\phi'_0 = 2 \arccos \frac{\omega}{a}\, .
\end{eqnarray}
Using this and applying (\ref{Lap}) to the intermediate result (\ref{intmed1}), we obtain
\begin{eqnarray}
I(\sigma) \sim \frac{2 \pi \sigma}{\sqrt{a^2 - \omega^2}} e^{\frac{2}{\sigma}\left( \sqrt{a^2 - \omega^2} - \omega \arccos \frac{\omega}{a}\right)}\, ,\quad \sigma \rightarrow 0\, . \label{intmed2}
\end{eqnarray}
The leading asymptotic behavior of $\bar{\omega}$ as $\sigma \rightarrow 0$ is then given by
\begin{eqnarray}
\bar{\omega}_{\rm asy} = \sqrt{a^2 - \omega^2} e^{- \frac{2}{\sigma}\left( \sqrt{a^2 - \omega^2} - \omega \arccos \frac{\omega}{a}\right)} \, . \label{omasy}
\end{eqnarray}
Figure \ref{fig1} shows the mean angular frequency $\bar{\omega}$ as a function of the noise level $\sigma$. The evaluation of the analytical expression (\ref{freqint}) yields results that are in good agreement with Monte-Carlo simulations of the Langevin equation (\ref{dyn}). Furthermore, the asymptotic expansion (\ref{omasy}) is in excellent agreement with numerical evaluations of (\ref{freqint}) for small noise.  

\section{Deterministic Two-Rotator System} \label{Sect4}
We next turn to a system of two coupled active rotators, where we first consider the deterministic case, i.e, $\sigma = 0$. In particular, we are interested in rotators with repulsive coupling, i.e., we consider the case $w_{12}, \, w_{21}>0$. Introducing the center of mass and difference coordinates $\Phi = (\phi_1 + \phi_2)/2$ and $\Delta = (\phi_1 - \phi_2)/2$, the set of equations (\ref{dyn}) takes the form
\begin{eqnarray}
\dot{\Phi}(t) &=& \omega - a \sin \Phi(t) \cos \Delta(t) \nonumber \\ 
	&& + (w_{12} - w_{21})\sin \Delta(t) \cos \Delta(t) \, , \nonumber \\
\dot{\Delta}(t) &=& - a \cos \Phi(t) \sin \Delta(t) \nonumber \\ 
&& + (w_{12}+w_{21}) \sin \Delta(t) \cos \Delta(t) \, . \label{2oscdet}
\end{eqnarray}
The system has a trivial stationary point at $\Phi(t) = \Phi_0 = \sin^{-1}(\omega/a)$, $\Delta(t) = 0$, whose stability we analyze by linearizing the system (\ref{2oscdet}). Writing $\Phi(t) = \Phi_0 + \epsilon_{\Phi}(t)$, $\Delta(t) = \epsilon_{\Delta}(t)$ we obtain to first order
\begin{figure*}[t]
  \begin{center}
    \epsfig{file=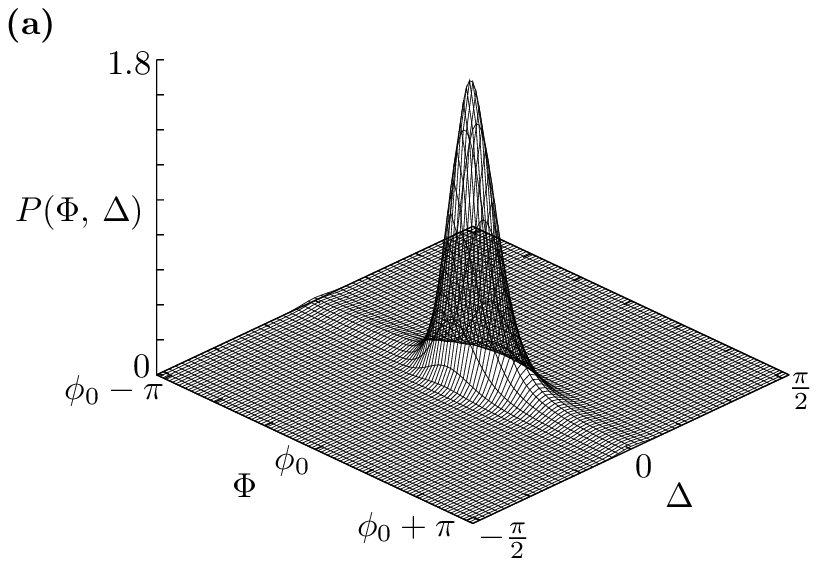,width=\columnwidth} 
    \hfill
    \epsfig{file=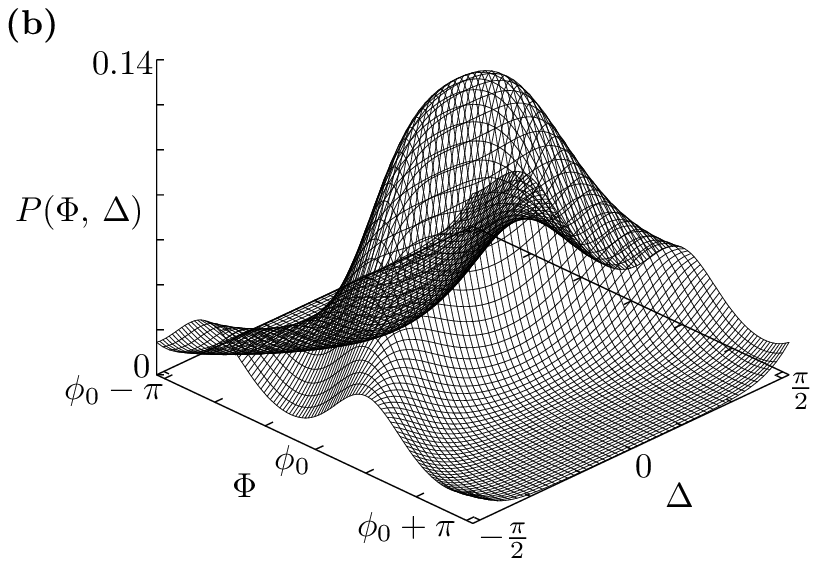,width=\columnwidth}
    \caption{Synchronized and desynchronized modes in the stochastic two-rotator system. The stationary solution to the Fokker-Planck equation (\ref{FPe}) is shown for different values of the coupling strength. In both {\bf (a)} and {\bf (b)}, we have $w_{12}=w_{21}=w$ and $\omega = 1$, $a = 1.2$, $\sigma = 0.4$. In {\bf (a)} the coupling strength is $w = 0.3$ and the rotators are in a synchronized state; in {\bf (b)} the coupling is increased to $w=0.4$ and the two rotators desynchronize.}
    \label{fig3}
  \end{center}
\end{figure*} 
\begin{eqnarray}
\frac{d}{dt} 
\left( \begin{array}{c}
\epsilon_{\Phi}(t) \\
\epsilon_{\Delta}(t)
\end{array}
\right)
&& \\
&\hspace{-40mm}=&\hspace{-20mm} 
\left( \begin{array}{cc}
-\sqrt{a^2 - \omega^2} & w_{12} - w_{21}  \\
0 & w_{12} + w_{21} - \sqrt{a^2 - \omega^2}  \end{array} \right) 
\left( \begin{array}{c}
\epsilon_{\Phi}(t) \\
\epsilon_{\Delta}(t)
\end{array}
\right)\, . \nonumber
\end{eqnarray}
The real parts of the eigenvalues of the $2 \times 2$ matrix on the right-hand side of the last identity determine the stability of the fixed point $(\Phi_0, 0)$. Under the assumption $a>\omega$ the first eigenvalue $\lambda_1 = - \sqrt{a^2 - \omega^2}$ is always real and negative.  The second eigenvalue $\lambda_2 = w_{12} + w_{21} - \sqrt{a^2 - \omega^2}$ is also always real; for small coupling it is negative, but when the sum of the coupling strengths $w_{12}+w_{21}$ increases it becomes positive and the fixed point $(\Phi_0, 0)$ loses its stability in, as it turns out, a subcritical pitchfork bifurcation. Further fixed points of the system can be determined and turn out to be unstable for all values of the coupling strengths. In the case $w_{12} = w_{21} = w$ they are given by
\begin{eqnarray}
\Phi_1 = \frac{1}{2} \sin^{-1}\left(\frac{4 \omega w}{a^2}\right), \quad
\Delta_1 = \cos^{-1}\left(\frac{\omega}{a \sin \Phi_1}\right).
\end{eqnarray}
Figure \ref{fig2}(a) shows a bifurcation diagram of the system. For small coupling strength, the system does not display oscillatory behavior. When the coupling strength is increased above a critical value, a stable limit cycle emerges from a homoclinic orbit. For a small range of coupling strengths, the stable fixed point coexists with the stable limit cycle. In this case, it depends on the initial conditions whether the system will converge toward the fixed point $(\Phi_0, 0)$ or the limit cycle. Figure \ref{fig2}(b) shows the attractors for fixed point and limit cycle dynamics in the $(\Phi,\Delta)$-plane for $w_{12} = w_{21} = 0.308$. In the strong-coupling limit, the minimum and maximum of $\Delta$ in Fig.~\ref{fig2}(a) both converge toward $\pi / 2$. Thus, the system approaches antisynchronous oscillatory dynamics, where $\phi_1$ and $\phi_2$ are phase shifted by $\pi$ while their sum increases constantly.  

\section{Stochastic Two-Rotator System} \label{Sect5}
We now consider the coupled two-rotator system in the case where both rotators receive uncorrelated stochastic driving. The temporal evolution of the probability density of this system is given by the Fokker-Planck equation (\ref{FPe}) with the drift and diffusion coefficients (\ref{drift}) and (\ref{diff}).
\subsection{Numerical Results} \label{Sect5a}
\begin{figure}[t]
  \begin{center}
    \epsfig{file=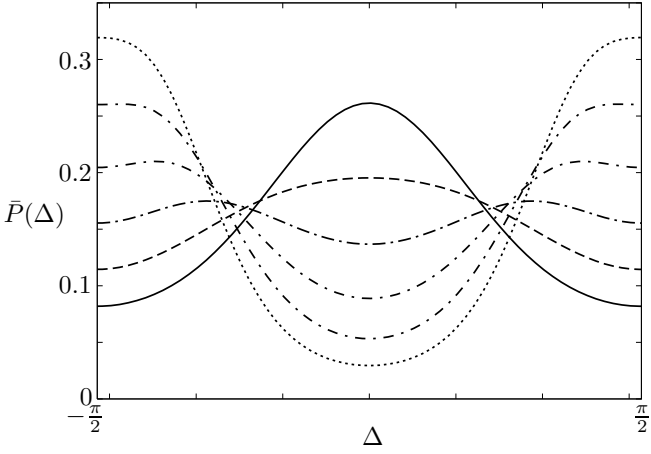,width=\columnwidth}    
    \caption{Marginal probability density for different values of the coupling strength $w = w_{12} = w_{21}$. The coupling strength for the curve with the highest value at $\Delta=0$ (solid line) is $w = 0.1$ and increases from curve to curve in increments of $\delta w = 0.2$ to the maximum value $w=1.1$ (dotted line). Other parameters are: $\omega = 1$, $\sigma = 0.4$, $a=1.2$.} 
    \label{fig4}
  \end{center}
\end{figure}

First, we investigate the stationary solution to the Fokker-Planck equation numerically. To this end, we numerically solve the partial differential equation (\ref{FPe}) under the periodic boundary conditions (\ref{perc}) for the homogeneous initial condition $P(\phi_1, \phi_2, t = 0) = 1 / 4 \pi^2$ and observe that the solution converges to the stationary solution after some time. Figure \ref{fig3} shows the stationary solution in the coordinates $\Phi$ and $\Delta$ for two different values of the coupling strength. We find that, depending on the strength of the noise and coupling, two different characteristic forms of the stationary solution exist. In the case shown in Fig.~\ref{fig3}(a) the probability density is peaked around the stable fixed point of the deterministic two-rotator system $(\Phi_0, 0)$. In Fig.~\ref{fig3}(b), the peak at the fixed point $(\Phi_0, 0)$ is much less pronounced. Furthermore, if we consider the probability distribution for $\Delta = \pm \pi/2$, i.e., at the edge of the region shown in Fig.~\ref{fig3}, we see that the probability distribution is not given by one central hump anymore. In order to distinguish between the two different scenarios in a quantitative way, we consider the marginal stationary probability density
\begin{eqnarray}
\bar{P}(\Delta) = \int_{\Phi_0 - \pi}^{\Phi_0 + \pi} d \Phi P(\Phi, \Delta) \,.  \label{Pmarg}
\end{eqnarray}
Figure \ref{fig4} shows this quantity for one level of the noise intensity $\sigma$ and for different coupling strengths. For weak coupling, $\bar{P}(\Delta)$ has a pronounced maximum at $\Delta = 0$. For increasing coupling strengths, this maximum decreases and eventually turns into a minimum. We can thus classify the system dynamics as synchronized or desynchronized according to the sign of the second derivative of $\bar{P}(\Delta)$ at the origin and can label the $\sigma$-$w$ plane accordingly. In the next section, we calculate the phase boundary between the synchronized and desynchronized regime through a Fourier expansion approach.

\subsection{Fourier Expansion Results} \label{Sect5b}
The probability density $P(\phi_1,\phi_2)$ is periodic in $\phi_1$ and $\phi_2$, so we expand it as
\begin{eqnarray}
P(\phi_1, \phi_2) = \sum_{k_1,k_2}C(k_1, k_2) e^{i(k_1 \phi_1 + k_2 \phi_2)}\, .
\end{eqnarray}
Inserting this approach into the right-hand side of (\ref{FPe}) yields together with (\ref{drift}) and (\ref{diff})
{\allowdisplaybreaks
\begin{eqnarray}
0 &=& \sum_{k_1, k_2} C(k_1, k_2)e^{i(k_1 \phi_1 + k_2 \phi_2)} \label{FourExp} \\
& \times & \Big\{ a (\cos \phi_1 + \cos \phi_2) - (w_{12} + w_{21}) \cos(\phi_2 - \phi_1) \nonumber \\ 
& - & i k_1 [ \omega - a \sin \phi_1 - w_{12} \sin(\phi_2 - \phi_1)] \nonumber \\
& - & i k_2[ \omega - a \sin \phi_2 - w_{21} \sin(\phi_1 - \phi_2)] - \sigma k_1^2 - \sigma k_2^2 \Big \}  \, . \nonumber
\end{eqnarray}
The term inside the curly brackets on the right-hand-side of the last identity is itself periodic in $\phi_1$ and $\phi_2$ and can also be expanded as a Fourier series
\begin{eqnarray}
\Big\{ \cdots \Big\} = \sum_{|l_1| \leq 1, |l_2| \leq 1} \tilde{C}(l_1, k_1, l_2, k_2) e^{i (l_1 \phi_1 + l_2 \phi_2)} \, .
\end{eqnarray}
Here, the coefficients $\tilde{C}(l_1, l_2)$ read
\begin{eqnarray}
\tilde{C}(0,k_1, 0, k_2) &=& - i \omega(k_1 + k_2) -  \sigma (k_1^2 + \sigma k_2^2)\,, \nonumber \\
\tilde{C}(\pm 1, k_1, 0, k_2) &=& \frac{a}{2} (1 \pm k_1)\, , \nonumber \\
\tilde{C}(0,k_1, \pm 1, k_2) &=& \frac{a}{2}(1 \pm k_2)\, ,  \nonumber \\
\tilde{C}(1,k_1,-1,k_2) &=& -\frac{1+k_1}{2}w_{12} - \frac{1 - k_2}{2} w_{21}\, , \nonumber \\
\tilde{C}(-1,k_1, 1, k_2) &=& -\frac{1-k_1}{2}w_{12} - \frac{1 + k_2}{2} w_{21}\,, \nonumber \\
\tilde{C}(\pm 1, k_1, \pm 1, k_2) &=& 0\,.
\end{eqnarray}
We can then rewrite (\ref{FourExp}) as
}
\begin{eqnarray}
0 & = & \sum_{k_1, k_2}e^{i(k_1 \phi_1 + k_2 \phi_2)} \\ 
&\times & \hspace{-2mm}\sum_{|l_1|<1, |l_2|<1}\hspace{-4mm} C(k_1 - l_1, k_2 - l_2) \nonumber  
\tilde{C}(l_1, k_1 - l_1, l_2, k_2 - l_2) \, .
\end{eqnarray}
Setting the inner sum to zero, we obtain an infinite system of algebraic equations. In order to obtain the $N$th Fourier order approximation we truncate the outer sum such that we set $C_N(k_1, k_2)=0$ for $|k_1| > N$ or $|k_2| > N$. Then, we have to solve a system of $(2N + 1)^2 - 1$ algebraic equations in order to obtain the expansion coefficients to $N$th order $C_N(k_1,k_2)$, where the additional index $N$ indicates the approximation order. Finally, the coefficient $C_N(0,0)$ is determined from the normalization condition as $C_N(0,0)=1/4\pi^2$. 

As an illustrative example we now consider the first order in the Fourier expansion for the case $w_{12}=w_{21}=w$. The system of algebraic equations we need to solve then reads
{\allowdisplaybreaks
\begin{eqnarray}
a[C_1(0,-1) + C_1(-1,0)] - 4(\sigma - i \omega)C_1(-1,-1)  &=& 0\, ,  \nonumber \\
4 \pi^2[2(\sigma - i \omega)C_1(-1,0) + w C_1(0, -1)] &=& a \, ,\nonumber \\
2 \pi^2 \{ a[C_1(-1, 0)+ C_1(0,1)]-4\sigma C_1(-1, 1)\} &=& w \, ,\nonumber \\
4 \pi^2[wC_1(-1,0) + 2(\sigma - i \omega)C_1(0, -1)] &=& a \, ,\nonumber \\
4 \pi^2[2 (\sigma + i \omega)C_1(0,1) + w C_1(1,0)] &=& a \, ,\nonumber \\
2 \pi^2 \{ a[C_1(0, -1) + C_1(1, 0)] - 4 \sigma C_1(1, -1)] \}  &=& w\, ,\nonumber \\
4 \pi^2[2(\sigma + i \omega)C_1(1,0) + w C_1(0, 1)]  &=& a\, ,\nonumber \\
a[C_1(0,1) + C_1(1,0)] - 4(\sigma + i \omega)C_1(1,1) &=& 0 . \nonumber \\
\end{eqnarray}
From this we obtain the first order approximation
\begin{eqnarray}
P_1(\phi_1,\phi_2) &=& \frac{1}{4\pi^2} +\alpha \{ 2 a \beta \gamma \sigma(\cos \phi_1 + \cos \phi_2) \nonumber \\ &+& 4a \beta \sigma \omega (\sin \phi_1 + \sin \phi_2) \nonumber \\
&+& (a^2 \gamma \sigma^2 - 2 a^2 \sigma \omega^2) \cos(\phi_1 + \phi_2) \nonumber \\ 
&+& [a^2 \beta \gamma - w \beta (\gamma^2 + 4 \omega^2)] \cos(\phi_1 - \phi_2) \nonumber \\
&+& a^2 \sigma( w \omega + 4 \sigma \omega) \sin (\phi_1 + \phi_2) \}\, ,
\end{eqnarray}
with the abbreviations
}
\begin{figure}[t]
  \begin{center}
    \epsfig{file=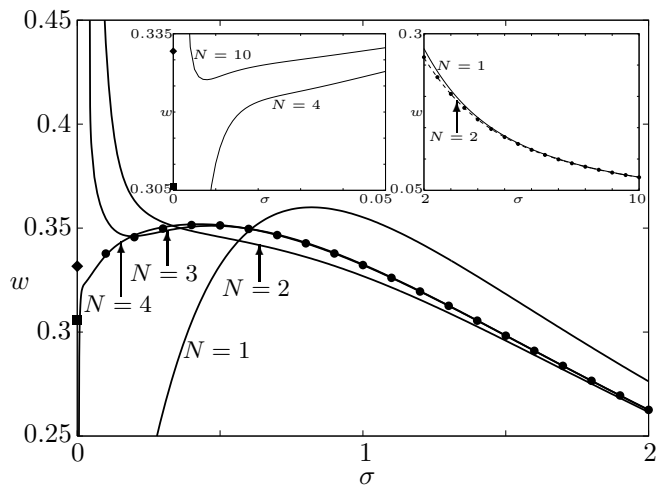,width=\columnwidth}    
    \caption{Regimes of synchronized and desynchronized dynamics. The phase boundary between the synchronized and desynchronized regimes is shown as a function of the noise strength $\sigma$. Areas below the curves correspond to the synchronized, areas above the curve to the desynchronized regime. The solid lines show the results of the first four Fourier orders, the dots represent numerical results. The diamond represents the coupling strengths for which the fixed point $(\Phi_0, 0)$ of the deterministic system becomes unstable; the square indicates the value of $w$ at which the stable limit cycle is first observed. The insets show results for small and for large noise. In the first inset (small noise) the results from the fourth and tenth Fourier orders are shown. The second inset (large noise) shows the results from the first (solid line) and second (dashed line) Fourier orders. Parameters are $\omega = 1$, $a=1.2$.}
    \label{fig5}
  \end{center}
\end{figure}
\begin{eqnarray}
\alpha = \frac{1}{4 \pi^2 \sigma \beta (\gamma^2 + 4 \omega^2)}\, , \quad \beta = \sigma^2 + \omega^2\, , \quad \gamma = w + 2 \sigma \, . \hspace{-2mm} \nonumber \\
\end{eqnarray}
Substituting the coordinates $\phi_1$ and $\phi_2$ according to $\phi_1 = \Phi + \Delta$ and $\phi_2 = \Phi - \Delta$ and integrating with respect to $\Phi$ we obtain the marginal probability density 
\begin{eqnarray}
\bar{P}_1(\Delta) = \frac{1}{2\pi} + 2 \pi \alpha \beta (a^2 \gamma - w \gamma^2 - 4 w \omega^2)\cos (2\Delta) \, .
\end{eqnarray}
Setting its second derivative to zero, we obtain the equation
\begin{eqnarray}
a^2(w + 2 \sigma) - w[(w + 2 \sigma)^2 + 4 \omega^2] = 0\, ,
\end{eqnarray}
which we can solve in $w$ or in $\sigma$. Eventually, we want to obtain $w$ as a function of $\sigma$. However, since we have a cubic equation in $w$ and only a quadratic equation in $\sigma$, for convenience we express $\sigma$ as a function of $\omega$:
\begin{eqnarray}
\sigma = \frac{a^2 - 2 w^2 \pm \sqrt{a^4 - 16w^2 \omega^2}}{4w}\,.
\end{eqnarray}
This procedure can easily be generalized to higher orders. Figure \ref{fig5} shows the resulting phase diagram obtained from solving the Fokker-Planck equation numerically and from the Fourier expansion. The accuracy of the Fourier expansion results improves with increasing strength of the noise. This can be seen, for instance, in the  second inset of Fig.~\ref{fig5}, where even the first expansion order yields very accurate results for strong noise. In general, even relatively low orders in the expansion give a good estimate for the phase boundary for a wide range of noise strengths, as can be seen from the results for the fourth expansion order in Fig.~\ref{fig5}. However, for very small noise levels the Fourier expansion diverges, as is exemplified in the first inset in Fig.~\ref{fig5} for the fourth and tenth expansion orders. Considering the first inset in Fig.~\ref{fig5}, we conclude that in the limit $\sigma \rightarrow 0$ the results from the Fourier expansion approach a value of the coupling strength for which the stable fixed point coexists with the limit cycle in the deterministic system. Therefore, neither the existence of the stable limit cycle nor the stability of the fixed point can be used exclusively to determine the zero-noise limit of the phase transition between the synchronized and desynchronized states. Strong noise has a desynchronizing effect on the system, as the minimal coupling for desynchronization vanishes in the limit of $\sigma \rightarrow \infty$. If the noise is weak, however, it stabilizes the synchronized state, as is indicated by the initially upward slope of the phase boundary in Fig.~\ref{fig5}. In conclusion, the synchronized state of the system is most stable for intermediate noise.

\section{Summary} \label{Sect6}
We have investigated the transition from synchronized to desynchronized behavior in a system of two-coupled active rotators under stochastic influences. The two regimes are distinguished by the sign of the second derivative of the marginal probability density at vanishing phase difference. We have evaluated the phase boundary between the two states in the (coupling strength) - (noise intensity) plane. Finally, we have shown that the synchronized state is most stable, in the sense that the coupling strength required to desynchronize the system is maximal for nonvanishing noise intensity.

\section{Acknowledgements}
We thank Janet Best, Anders Carlsson, John Clark, and John Rinzel for fruitful discussions. This work was supported in part by NIH-EY 15678. 


\end{document}